\documentclass[twocolumn,preprintnumbers,nofootinbib,prd,superscriptaddress,aps]{revtex4-1}

\usepackage[utf8]{inputenc} 
\usepackage{graphicx,amssymb,amsmath,amsthm,amsfonts,epstopdf,epsfig,epsf,times}
\usepackage[linktocpage]{hyperref}
\usepackage[usenames]{color}
\usepackage{epstopdf}
\usepackage{textcomp}
\usepackage{bm}
\usepackage{latexsym}
\usepackage{rotating}
\usepackage{hyperref}
\usepackage{color}
\usepackage{longtable}
\usepackage{enumerate}
\usepackage{tensor}
\usepackage{stmaryrd}
\usepackage[normalem]{ulem}
\usepackage{mathtools}
\usepackage{url}
\usepackage{multirow}
\usepackage{graphicx}
\usepackage{mathtools}
\usepackage{verbatim}
\usepackage{soul,xcolor}
\setstcolor{red}

\definecolor{coolblack}{rgb}{0.0, 0.18, 0.39}
\definecolor{darkred}{rgb}{0.5,0,0}
\definecolor{darkgreen}{rgb}{0,0.5,0}
\definecolor{darkblue}{rgb}{0,0,0.5}
\definecolor{lapislazuli}{rgb}{0.15, 0.38, 0.61}
\definecolor{venetianred}{rgb}{0.78, 0.03, 0.08}
\definecolor{bleudefrance}{rgb}{0.19, 0.55, 0.91}
\definecolor{dogwoodrose}{rgb}{0.84, 0.09, 0.41}
\definecolor{dogwoodrose}{rgb}{0.84, 0.09, 0.41}
\definecolor{darkorgane}{rgb}{1,0.549,0}
\hypersetup{colorlinks=true, citecolor=darkblue, linkcolor=darkblue, 
urlcolor = darkblue}
\definecolor{olive}{rgb}{0.5, 0.5, 0.0}

\newcommand{\ben}{\begin{enumerate}}
\newcommand{\een}{\end{enumerate}}

\newcommand{\del}{\partial}

\def\be{\begin{equation}}
\def\ee{\end{equation}}
\newcommand{\beq}{\begin{eqnarray}}
\newcommand{\eeq}{\end{eqnarray}} 
\newcommand{\ba}{\begin{align}}
\newcommand{\ea}{\end{align}}

\def\be{\begin{equation}}
\def\ee{\end{equation}}
	
\newcommand{\bea}{\begin{eqnarray}}
\newcommand{\eea}{\end{eqnarray}}

\begin{document}
\title{Black holes in galaxies: environmental impact on gravitational-wave generation and propagation}

\author{Vitor Cardoso}
\affiliation{Niels Bohr International Academy, Niels Bohr Institute, Blegdamsvej 17, 2100 Copenhagen, Denmark}
\affiliation{CENTRA, Departamento de F\'{\i}sica, Instituto Superior T\'ecnico -- IST, Universidade de Lisboa -- UL,
Avenida Rovisco Pais 1, 1049 Lisboa, Portugal}
\author{Kyriakos Destounis}
\affiliation{Theoretical Astrophysics, IAAT, University of T{\"u}bingen, 72076 T{\"u}bingen, Germany} 
\author{Francisco Duque}
\affiliation{CENTRA, Departamento de F\'{\i}sica, Instituto Superior T\'ecnico -- IST, Universidade de Lisboa -- UL,
Avenida Rovisco Pais 1, 1049 Lisboa, Portugal}
\author{Rodrigo Panosso Macedo}
\affiliation{CENTRA, Departamento de F\'{\i}sica, Instituto Superior T\'ecnico -- IST, Universidade de Lisboa -- UL,
Avenida Rovisco Pais 1, 1049 Lisboa, Portugal}
\author{Andrea Maselli}
\affiliation{Gran Sasso Science Institute (GSSI), I-67100 L’Aquila, Italy}
\affiliation{INFN, Laboratori Nazionali del Gran Sasso, I-67100 Assergi, Italy} 

\begin{abstract} 
We introduce a family of solutions of Einstein's gravity minimally coupled to an anisotropic fluid, describing asymptotically flat black holes with ``hair'' and a regular horizon. These spacetimes can describe the geometry of galaxies harboring supermassive black holes, and are extensions of Einstein clusters to include horizons. They are useful to constrain the environment surrounding astrophysical black holes, using electromagnetic or gravitational-wave observations. We compute the main properties of the geometry, including the corrections to the ringdown stage induced by the external matter and fluxes by orbiting particles. The leading order effect to these corrections is a gravitational-redshift, but gravitational-wave propagation is affected by the galactic potential in a nontrivial way, and may be characterized with future observatories.
\end{abstract}
\maketitle


\noindent{\bf{\em Introduction.}}
There is overwhelming evidence for the existence of dark matter of an unknown nature, but which interacts gravitationally~\cite{Freese:2008cz,Navarro:1995iw,Clowe:2006eq,Bertone:2004pz}.
Understanding the properties of dark matter and its context in the ``Standard Model'' of particle physics is arguably one of the outstanding challenges in science.
Efforts to detect possible minute interactions between the dark sector and the standard model have so far been unsuccessful, but will continue vigorously for years to come~\cite{Kahlhoefer:2017dnp,PerezdelosHeros:2020qyt}.

In the meanwhile, the advent of gravitational-wave (GW) astronomy~\cite{LIGOScientific:2016aoc,Abbott:2020niy} and of very-long baseline interferometry~\cite{EventHorizonTelescope:2019dse,GRAVITY:2020gka} may revolutionize our ability to study the invisible universe~\cite{Barack:2018yly,Cardoso:2019rvt,Bertone:2018krk,Bar:2019pnz,Brito:2015oca}. These observatories are tailored to study compact objects, such as black holes (BHs). Dark matter may cluster at the center of galaxies and close to BHs~\cite{Sadeghian:2013laa}, affecting the dynamics of compact binaries and the way in which GWs or electromagnetic waves propagate~\cite{Eda:2013gg,Macedo:2013qea,Barausse:2014tra,Barack:2018yly,Baibhav:2019rsa,Seoane:2021kkk}.

How {\it does} surrounding matter affect the generation and propagation of GWs or the electromagnetic appearance of BHs?
To answer this question, a motivated matter distribution and corresponding spacetime geometry must be known.
So far, however, studies have focused mostly on Newtonian approaches to the problem: the self-gravity of the matter surrounding a BH was modeled at a Newtonian level,
and its impact on the dynamics of compact binaries was quantified by adding estimates of dynamical friction~\cite{Eda:2013gg,Macedo:2013qea,Barausse:2014tra,Cardoso:2019rou,Kavanagh:2020cfn}.
Possible gravitational redshift or peculiar motion effects have also been modeled with resort to Doppler-like changes to waveforms~\cite{Tamanini:2019usx}, but without a first-principles calculation.

Clearly, to go beyond crude estimates, or to quantify the impact of an entire galaxy on the propagation of GWs, one needs to model the geometry.
A good description of galactic profiles is guided by observations and large-scale simulations.
The profiles that are observed in bulges and elliptical galaxies are well described by a Hernquist-type density distribution~\cite{1990ApJ...356..359H}
\be
\rho=\frac{Ma_0}{2\pi\,r(r + a_0)^3}\,,\label{eq_rho:hernquist}
\ee
where $M$ is the total mass of the ``halo'' and $a_0$ a typical lengthscale. There are other popular density profiles -- namely Navarro-Frenk-White~\cite{Navarro:1995iw}, Jaffe~\cite{1983MNRAS.202..995J} or King~\cite{1962AJ.....67..471K} profiles -- adjusting well to dark matter simulations.

Note that the Hernquist profile -- as well as others in the same ``family''~\cite{1990ApJ...356..359H,Navarro:1995iw,1983MNRAS.202..995J,1962AJ.....67..471K} --
have an increased density in the cores of the galaxies. However, in the presence of a BH at the core, Newtonian and relativistic analysis show that the density profile
is zero close to the horizon~\cite{Gondolo:1999ef,Sadeghian:2013laa}. The dark matter profile develops a cusp close to the horizon, with a lengthscale dictated by the BH mass $M_{\rm BH}$ (therefore the total mass in this cusp is negligibly small). The density profile vanishes at the horizon. The precise details of the profile depend on the equation of state, but eventually give way to the larger scale family of Hernquist or Hernquist-like profiles above.

An important parameter determining the gravitational properties of a galaxy is the
compactness of the ``halo'' $GM/(c^2a_0)$ which also determines the virial velocity. We will take this parameter to be as large as $10^{-4}$ for galaxies~\cite{Navarro:1995iw}, but in the context of constraints on BH environments, the parameter is basically free. Here onwards we set Newton's constant $G$ and the speed of light $c$ to unity.

\noindent{\bf{\em The geometry of black holes in galactic centers.}}
We now wish to place a BH at the center of distribution~\eqref{eq_rho:hernquist}. At a Newtonian level, the procedure is trivial~\cite{1996ApJ...471...68C,2004MNRAS.351...18B}.
We wish to find a general-relativistic geometry which on small scales describes a BH and on large scales describes matter distributed according to~\eqref{eq_rho:hernquist}. For simplicity, we assume that the system is spherically symmetric. 
We can follow Einstein in his construction of a stationary system of many gravitating masses, an ``Einstein cluster''~\cite{Einstein:1939ms,Geralico:2012jt}, and generalize it to include a central BH. The procedure effectively takes particles in all possible circular geodesics, and deals with an ``average'' stress-energy tensor~\cite{Einstein:1939ms,Geralico:2012jt}, characterized by the matter density $\rho$. 

The Einstein construction assumes a stress tensor $\langle T^{\mu\nu}\rangle=\frac{n}{m_p}\langle P^{\mu}P^{\nu}\rangle$, with $n$ the number density of particles with mass $m_p$ and $P$ the four-momentum satisfying the geodesic equations. Alternatively, it is easily shown that this construction is equivalent to assuming an anisotropic material with only tangential pressure $P_t$, and vanishing radial pressure, 
\be
T^\mu_\nu={\rm diag}(-\rho,0,P_t,P_t)\,.
\ee 
In particular, if one writes the geometry as
\be
ds^2=-fdt^2+\frac{dr^2}{1-2m(r)/r}+r^2d\Omega^2\,,
\label{eq:Spherical}
\ee
then the radial function $f$ is obtained from the mass function $m(r)$ (which we take to be provided by observations) as
\be
\frac{rf'}{2f}=\frac{m(r)}{r-2m(r)}\,,\label{eq_metric}
\ee
while the tangential pressure can be obtained from the Bianchi identities
\be
2P_t=\frac{m(r)\rho}{r-2m(r)}\,.
\ee

Here, we explore one possible matter distribution inspired by the Hernquist profile~\eqref{eq_rho:hernquist}. In particular, we assign
\be
m(r)=M_{\rm BH}+\frac{M r^2}{(a_0+r)^2}\left(1-\frac{2M_{\rm BH}}{r}\right)^2\,,
\ee
and require asymptotic flatness (i.e., $f\to 1$ at large $r$). At small distances this profile describes a source of mass $M_{\rm BH}$. At large distances,
the density profile corresponds to the distribution~\eqref{eq_rho:hernquist}.
It turns out that the profile above has a very simple BH solution for the metric components, which we can explore to understand the phenomenology of BHs and GWs from
objects deep in the galactic potential. The solution is
\beq
%
f&=&\left(1-\frac{2M_{\rm BH}}{r}\right)e^\Upsilon\,,\label{eq_fhairy}\\
\Upsilon&=&-\pi\sqrt{\frac{M}{\xi}}+2\sqrt{\frac{M}{\xi}}\arctan{\frac{r+a_0-M}{\sqrt{M\xi}}}\,,\\
\xi&=&2a_0-M+4M_{\rm BH}\,.
%
\eeq
%
To mimic observations of galaxies, $a_0 \gtrsim 10^{4}M$. The solution above corresponds to a matter density
\be
4\pi \rho=\frac{m'}{r^2}=\frac{2M(a_0+2M_{\rm BH})(1-2M_{\rm BH}/r)}{r(r+a_0)^3}\,.\label{eq:hernquist_GR}
\ee
%
Some properties of this geometry are worth highlighting.
The first is that it is a BH spacetime indeed, with an horizon at $r=2M_{\rm BH}$ and a curvature singularity at $r=0$.
At large distances, the Newtonian potential is that of the Hernquist profile and dominated by a ``halo'' of mass $M$. The ADM mass of the spacetime is $M+M_{\rm BH}$.
The matter density vanishes at the horizon and the tangential pressure is regular there. 
%
When there is a hierarchy of scales, i.e. $M_{\rm BH}\ll M\ll a_0$, then the Ricci scalar $R\sim M/(a_0^2 M_{\rm BH})$ close to the horizon, which can be made small in a controlled way. At large distances, it decays as $\sim 4Ma_0/r^4$. 
For parameters which do not describe astrophysical setups, singularities may occur. In particular, for $M>2(a_0+2M_{\rm BH})$ there are curvature singularities also at $r=M-a_0 \pm\sqrt{M^2-2Ma_0-4MM_{\rm BH}}$ where the Ricci and Kretschmann scalars diverge.
%

The anisotropic fluid ``hair'' surrounding the BH satisfies the weak and strong energy conditions everywhere, since both pressure and density are always positive.
The light ring is located at the root of $r=3m(r)$~\cite{chandrasekhar1992mathematical,Cardoso:2008bp}, where the tangential pressure satisfies $P_t=\rho/2$ for {\it any} mass function. At 
$r=2m(r)$ there is an horizon and $P_t/\rho$ diverges. Thus, the dominant energy condition is violated close to the horizon.
The near-horizon region is, as we said, nearly empty: even though the dominant energy condition is violated, both the pressure and density are arbitrarily small in this region,
and thus will play no role in the spacetime dynamics.

The redshift factor $e^\Upsilon\sim 1-2M/a_0$ close to the central BH. We find this to be a property of BH geometries with
profiles other than~\eqref{eq:hernquist_GR} as well (we explored in particular Plummer-like~\cite{1911MNRAS..71..460P} and boson star profiles~~\cite{Liebling:2012fv,Annulli:2020lyc}). 

\noindent{\bf{\em Redshift, light rings and ISCOs.}}
For generic mass profiles the light ring is displaced from its pure vacuum location, and lies at
(in describing geodesics we keep terms up to ${\cal O}(1/a_0^3)$)
\be
r_\textnormal{LR}\approx 3M_{\rm BH}\left(1+\frac{MM_{\rm BH}}{a_0^2}\right)\,,
\ee
with the second term a minute fraction of the first. On the other hand, the light ring frequency $\Omega_{\rm LR}$ and Lyapunov scale $\lambda$~\cite{Cardoso:2008bp} are redshifted to
\beq
&&M_{\rm BH}\Omega_{\rm LR}\sim \frac{1}{3\sqrt{3}}\left(1-\frac{M}{a_0}+\frac{M(M+18M_{\rm BH})}{6a_0^2}\right)\,,\label{eq:lr_redshift}\\
&&M_{\rm BH}\lambda\sim \frac{1}{3\sqrt{3}}\left(1-\frac{M}{a_0}+\frac{M^2}{6a_0^2}\right)\,.\label{eq:lr_lyapunov}
\eeq
For galaxy descriptions, the first two terms dominate and are simply a (Newtonian) gravitational redshift effect. The geometry admits stable light rings 
for certain parameters. The critical impact parameter $b_{\rm crit}$ for capture of high-frequency photons or GWs is then~\cite{chandrasekhar1992mathematical}
\be
b_{\rm crit}=3\sqrt{3}M_{\rm BH}\left(1+\frac{M}{a_0}+\frac{M(5M-18M_{\rm BH})}{6a_0^2}\right)\,.
\ee
Thus, at dominant order, haloes of mass $M$ only redshift the dynamics. Our results show that the galactic content
affects the shadow of BHs by terms of order $M^2/a_0^2\lesssim 10^{-8}$. In other words, tests of the nature of the central object using light-ring physics can still be done, with the Event Horizon Telescope, GRAVITY or similar instruments~\cite{EventHorizonTelescope:2019dse,2018A&A...618L..10G}, to a good precision.

There are stable timelike circular geodesics of redshifted energy per unit rest mass $E^2=2f^2/(2f-rf')\sim E^2_{\rm vacuum}(1-2M/a_0)$~\cite{Cardoso:2008bp} for $r>r_\textnormal{ISCO}$. 
We find that the innermost stable circular orbit (ISCO) is also redshifted to
\beq
r_\textnormal{ISCO}\sim 6M_\textnormal{BH}\left(1-\frac{32MM_\textnormal{BH}}{a_0^2}\right)\ ,
\eeq
and the corresponding angular frequency is 
\beq
M_{\rm BH}\Omega_{\rm ISCO}\sim \frac{1}{6\sqrt{6}}\left(1-\frac{M}{a_0}+\frac{M(M+396M_{\rm BH})}{6a_0^2}\right)\,. \nonumber \\\label{isco}
\eeq
The non vacuum background also induces a shift in the vertical and radial  
oscillation frequencies~\cite{Maselli:2014fca}. At the ISCO, $\Omega_r=0$ while, coincidentally,
$\Omega_\theta=\Omega_{\rm ISCO}$.

\noindent{\bf{\em Love numbers, quasinormal modes and tails.}}
%
\begin{figure}[ht!]
\begin{tabular}{c}
\includegraphics[scale=0.39]{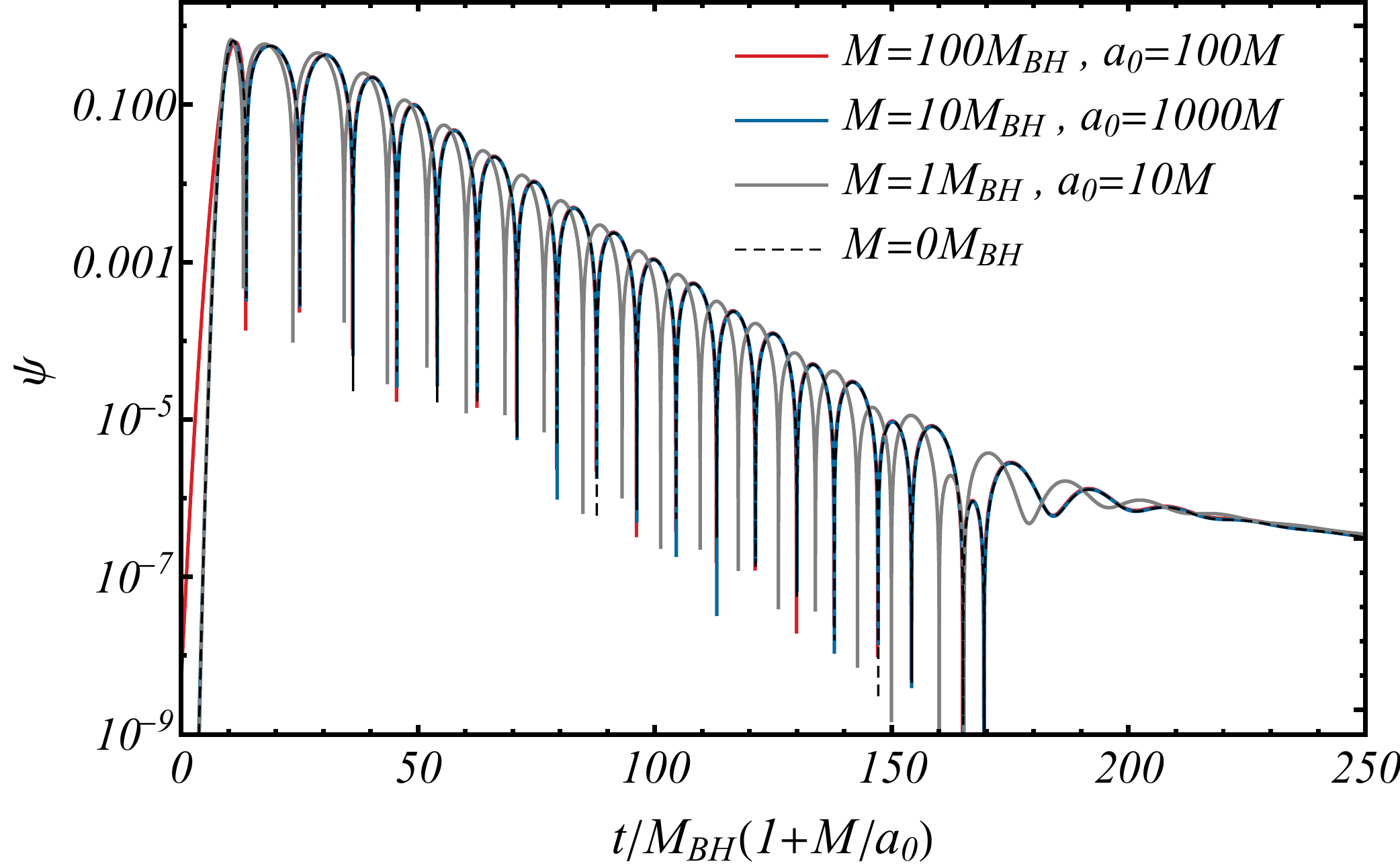}\\
\includegraphics[scale=0.39]{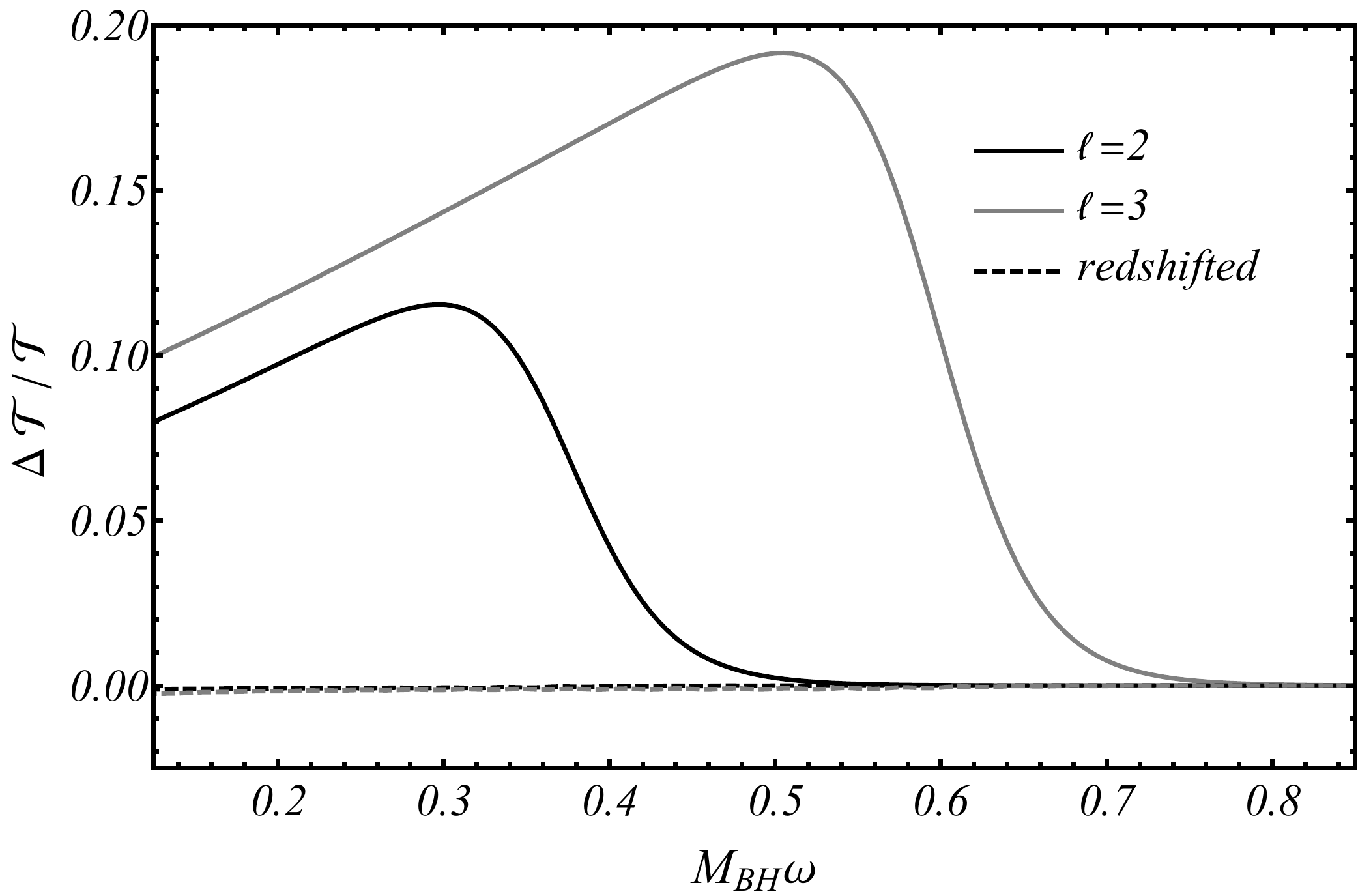}
\end{tabular}{}
\caption{\textit{Top panel}: time evolution of a Gaussian wavepacket $\del_t \psi (t=0, r_*) = \exp\left[ - (r_* - 10)^2  \right]$ of $\ell=2$ axial-type GWs, in the spacetime of a galaxy with a BH at the center, geometry~\eqref{eq_fhairy}, for different ``halo'' properties. The signal is extracted at null infinity. Waveforms were aligned and redshifted by the factor $1-M/a_0$. 
\textit{Bottom panel}: a scattering experiment for $M=10 M_{\rm BH}$ and $a_0=100M$: a monochromatic pulse of GWs is thrown into the BH, a fraction ${\cal T}$ of the flux is absorbed by the horizon. This plot shows the relative difference in the transmittance ${\cal T}$, in the presence or absence of a halo (i.e., with $M=0$). The dashed lines show the corresponding quantity when the pulse incident on a BH with $M\neq 0$ is redshifted. These results indicate once again that as the GWs fall and get close to the central BH they are blueshifted, therefore giving rise to the same transmittance as that of a vacuum BH. \label{axial_scatters}}
\end{figure}
The geometry~\eqref{eq_fhairy} changes GW generation and propagation, with respect to a vacuum BH. To study the possible effects on GW propagation, we now consider spacetime fluctuations, decomposing the perturbed metric in tensor harmonics~\cite{Regge:1957td,Zerilli:1970wzz,Maggiore:2007ulw}.
For axial-type perturbations, the only non-vanishing components of the metric fluctuations are
$h_{t\phi}=h_0\sin\theta Y_{\ell m}'(\theta)e^{-i\omega t}$ and $h_{r\phi}=h_1\sin\theta Y_{\ell m}'(\theta)e^{-i\omega t}$, where $h_0$ and $h_1$ are functions of $t$ and $r$. It is straightforward to combine them in a master wave equation
~\cite{1967ApJ...149..591T,1983ApJS...53...73L} for the variable $\psi^\textnormal{(ax)}$ governed by
\beq
&&\frac{d^2\psi^\textnormal{(ax)}}{dr_\star^2}+\left[\omega^2-V^\textnormal{(ax)}\right]\psi^\textnormal{(ax)}=S^\textnormal{(ax)}\,,\\
&&V^\textnormal{(ax)}=f\left(\frac{\ell(\ell+1)}{r^2}-\frac{6m}{r^3}+\frac{m'}{r^2}\right)\,.\label{eq:Vax}
\eeq
Here the tortoise coordinate is defined as $dr/dr_\star=\sqrt{f(1-2m/r)}$ and the source term $S^\textnormal{(ax)}$ describes putative distributions of matter generating gravitational perturbations.
In the large $a_0$ limit, $V^\textnormal{(ax)}=V^\textnormal{(ax)}_{\rm vacuum}(1-2M/a_0)$ and the master wave equation reduces to vacuum
if the frequency is redshifted to $\omega^2 (1-2M/a_0)$. This might be anticipated and it is pleasant to see it arising from the formalism.

Using the above decomposition of the metric tensor, one can compute the tidal Love numbers of these BH solutions~\cite{Binnington:2009bb,Damour:2009vw,Cardoso:2017cfl}. A closed-form analytical expression can be found at small $M$, which also describes well our full numerical results at large $M$, 
\be
k_{\ell=2}^B=\frac{Ma_0^4(5+12\log{(a_0+2M_{\rm BH})})}{3(M+M_{\rm BH})^5}\,.
\ee
The scaling with $M$ and $a_0$ might be anticipated from previous discussions in the literature~\cite{Cardoso:2019upw}. For possible caveats when using these numbers to understand dynamics of extended objects see Ref.~\cite{Cardoso:2019upw}.

We have studied the characteristic quasinormal frequencies of the problem \eqref{eq:Vax}, using standard direct integration and spectral routines~\cite{Berti:2009kk,Jansen:2017oag,Cardoso:2017soq}.
These find a discrete set of complex frequencies $\omega_{\ell,n}(M,\,a_0)$ satisfying appropriate outgoing boundary conditions.
We find that the behavior of the quasinormal modes is consistent with a light ring interpretation~\cite{Cardoso:2008bp}. For $M\ll M_{\rm BH}$ but large $a_0\gg M_{\rm BH}$ a fit of our numerical results to powers of $M/a_0$ yields, for the fundamental quadrupolar mode,
\be
\frac{\omega_{20}(M,a_0)}{\omega_{20}(0,0)}\sim 1-1.1\frac{M}{a_0}+{\cal O}\left(\frac{M^2}{a_0^2}\right)\,,\label{eq:ringdown}
\ee
for both the real and imaginary component separately, with errors $\lesssim 3\%$. This result is consistent with the properties \eqref{eq:lr_redshift}.
Again, the leading order correction is due to gravitational redshift, but there are subdominant contributions whose detection is challenging if $M/a_0\lesssim 10^{-4}$~\cite{Berti:2005ys,Berti:2016lat}.

To understand the free evolution of a BH in this geometry, we repeated Vishveshwara's classical scattering experiment, 
letting a gaussian wavepacket evolve in this background~\cite{Vishveshwara:1970zz}. Our results are summarized in Fig.~\ref{axial_scatters}. 
The scattering excites the characteristic vibration modes of the BH~\cite{Berti:2009kk,Cardoso:2016rao,Cardoso:2016oxy,Cardoso:2019rvt}, and the signal damps exponentially. At late times, backscattering off spacetime curvature gives rise to a power-law tail~\cite{Price:1971fb,Ching:1994bd,Blanchet:1994ez}.
Our results show no evidence for anomalous ringdown, of the type encountered in late-time echoes~\cite{Cardoso:2016rao,Cardoso:2016oxy,Cardoso:2017cqb,Cardoso:2019rvt}, which would 
signal a scale other than that set by the BH, $M_{\rm BH}$. This contrasts with previous models of galaxy matter, where it was observed that a simple shell far away from the BH would produce light-ring relaxation, followed by quasinormal ringdown~\cite{Barausse:2014tra,Cardoso:2019rvt}. The simplicity of the relaxation of geometry~\eqref{eq_fhairy} relates to a smoother matter distribution, and to absence of trapping regions outside the light ring~\cite{Zworski99,SjoZwo99}.
Instead, our results show that the ringdown is consistent with a quasinormal mode analysis, which in turn yields frequencies which are, to dominant order, redshifted relative to the vacuum as in Eq.~\eqref{eq:ringdown}. In other words, the exponentially decaying stage of this signal can be interpreted as GWs leaking from the light ring (cf. Eq.~\eqref{eq:lr_redshift}) and escaping to infinity while being redshifted due to their climbing of the gravitational potential. Exceptions to this interpretation occur only when $M/a_0$ is large, implying that the near-horizon region is changed by the surrounding matter. Our time evolutions are consistent with quasinormal mode calculations even in this regime.

On the other hand, one could expect late-time tails to be radically different since they are low-frequency phenomena (in fact zero-frequency~\cite{Price:1971fb,Ching:1994bd,Blanchet:1994ez}) and therefore able to probe the geometry arbitrarily far away from the BH. Our results, summarized in the top panel of Fig.~\ref{axial_scatters}, show otherwise. For pulses released close to the horizon, 
the signal is simply redshifted, to dominant order.

\noindent{\bf{\em Scattering and energy emission from orbiting bodies.}}
\begin{figure}[ht!]
\includegraphics[scale=0.55]{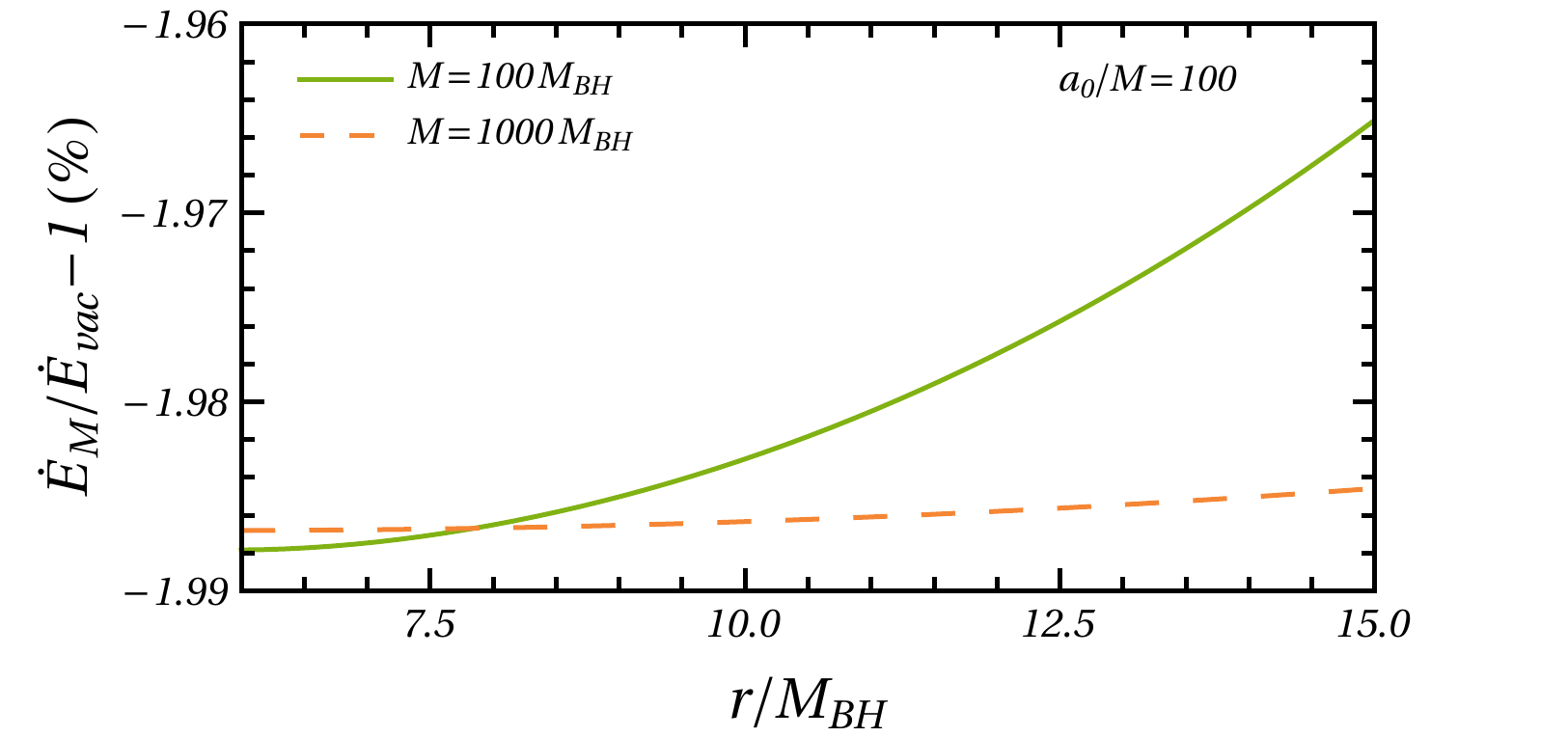}
\caption{Relative percentage difference between the $(\ell,m)=(2,1)$  
axial component of the flux computed in vacuum and for different 
choices of $(M,a_0)$ as a function of the orbital radius of 
the secondary BH.
\label{axial_circular}}
\end{figure}
We now turn to stimulated GW emission. One can start probing the spacetime by scattering a monochromatic pulse of radiation, such that asymptotically at spatial infinity
\be
\psi^\textnormal{(ax)}={\cal I} e^{-i\omega r}+{\cal R} e^{i\omega r}\,,\qquad r\to \infty \, .\\
\ee
Define the transmittance 
\be
{\cal T}(M,\omega)\equiv 1-\frac{|{\cal R}|^2}{|{\cal I}|^2}\,,
\ee
which is a measure of the fraction of incident flux crossing the horizon. This function is well studied in vacuum geometries~\cite{Teukolsky:1974yv,Brito:2015oca}.
To quantify the impact of the surrounding matter, define $\Delta {\cal T}\equiv 1-{\cal T}(0,\omega)/{\cal T}(M,\omega)$. Results are shown in Fig.~\ref{axial_scatters} (bottom panel), for $M=10 M_{\rm BH},\, a_0=100M$. Differences are of order $10\%$ close to the quadrupolar resonant frequency of the BH, and higher for higher modes. Results are weakly sensitive to $M$, but depend mostly on $M/a_0$ only. We also show the redshifted version $\Delta \bar{\cal T}\equiv 1-{\cal T}(0,\omega)/{\cal T}(M,\omega(1-M/a_0))$, which is extremely small. Our results are consistent with a redshift interpretation:  when thrown from infinity, a GW of lower frequency gets blueshifted and interacts with the BH in the same way that it would in vacuum. The transmittance is mostly decided by the geometry close to the horizon, hence $\Delta \bar{\cal T}\sim 0$. Note a slight disagreement at smaller frequencies, lending further support to this interpretation: high frequencies are simply redshifted. Low frequencies probe the bulk of the spacetime and are therefore affected beyond such leading order. The previous picture holds true for large $a_0$ and small $M/a_0$ for which the near horizon region is nearly vacuum. 

There are low-frequency corrections. For $\ell=2$ and $10^{-3} \lesssim  M_{\rm BH} \omega\lesssim 0.2$, we find
\be
\Delta \bar{\cal T}\approx
10^{-9}\frac{M}{10^4M_{\rm BH}}\frac{0.01}{M_{\rm BH}\omega}\left(\frac{10^4M}{a_0}\right)^2\,.
\ee
This effect is small, but not desperately small. The impact of such correction on the phasing of GW signals may be large enough to be detectable by third generation and space-based GW detectors~\cite{LISA:2017pwj,Seoane:2021kkk,Maggiore:2019uih}.

To study a process directly of relevance to GW astronomy, consider now 
a small point particle (representing for example a neutron star or a stellar-mass BH) orbiting the BH spacetime close to the ISCO, 
which induces a non-trivial source $S^{\rm (ax)}$~\cite{Zerilli:1970wzz,Martel:2005ir}. 
The relative difference in the energy flux in the axial sector is shown in Fig.~\ref{axial_circular} for $a_0=100M$. 
In general, the discrepancy between the two GW fluxes grows with $M$. The difference is suppressed as the ratio $M/a_0$ decreases, and 
it becomes $\ll 1\%$ for $M/a_0=10^{-3}$. Given the values of $\dot{E}$ we 
can quantify the impact of the halo configurations on possible detections 
of extreme-mass-ratio inspirals (EMRIs) by the future satellite LISA~\cite{LISA:2017pwj}. 
To this aim we determine the orbital phase evolution, assuming that the polar sector is shifted form the vacuum 
by the same fractional amount as the axial. We take a prototype binary 
with component masses $M_{\rm BH}=10^6M_\odot$ and $m_2=10M_\odot$, observed 
for one year before the ISCO~\cite{Maselli:2020zgv}. We compute the total 
number of cycles ${\cal N}$ of such binaries, in vacuum or in presence of a halo. 
For the two cases shown in Fig.~\ref{axial_circular} we obtain 
${\cal N}_M-{\cal N}_{vac}\sim 500$. 
Clearly, a more sophisticated analysis is required~\cite{Maselli:2021men}, but such dephasing 
provides good indications that the halo may affect GW generation and propagation at observable levels.

\noindent{\bf{\em Discussion.}}
The spacetime described by Eq.~\eqref{eq_fhairy} is a simple geometry describing a BH within a galactic core of matter. Such spacetime is an interesting starting point to understand GW generation and propagation deep in galactic cores, or to test more generically the geometry of BH spacetimes (in which case the parameters $(M,\,a_0)$ do not necessarily take on the values expected for galaxies).
A more realistic solution should be totally depleted inside the light ring, though possibly not inside the ISCO~\cite{Sadeghian:2013laa}.
To test the effect of such distributions, we integrated Eq.~\eqref{eq_metric} numerically for mass profiles truncated at the ISCO or close to it (and replacing the $1-2M_{\rm BH}/r$ factor with some other factor, for example $1-6M_{\rm BH}/r$). All relevant energy conditions are then satisfied everywhere, and we find that some of the important properties of the geometry remain basically the same, including the light ring frequency down to factors ${\cal O}(1/a_0^3)$, as shown in Eq.~\eqref{eq:lr_redshift}. 
Similar constructions can be made for any density profile, with or without a cosmological constant. However, simple closed-form and realistic solutions are challenging to find. 
We note, for example, that recent attempts at describing such geometries used a simplistic version of Einstein's procedure and assumed $f=1-2m(r)/r$. The resulting geometries have 
radial pressure and anomalous behavior close to the event horizon and at spatial infinity~\cite{Xu:2018wow}. See also Ref.~\cite{Cadoni:2017evg} for construction of spacetimes sourced by anisotropic fluids.

We note that previous studies have considered Hernquist-like profiles with central pointlike particles in a Newtonian but not general-relativistic setting~\cite{1996ApJ...471...68C,2004MNRAS.351...18B}. Stability analysis of such solutions were performed via N-body simulations (see e.g. Ref.~\cite{Buyle:2006vs}). A fully relativistic analysis of radial perturbations of the geometry~\eqref{eq_fhairy} can be done following Chandrasekhar~\cite{Chandrasekhar:1964zz,1964ApJ...140..417C,1966ApJ...145..505B,Kokkotas:2000up,Dev:2003qd}, extended to include anisotropic pressure.
It indicates marginal stability for some equations of state, although we do expect dynamical instability in some (possibly non-radial) modes, due to matter content inside the light ring. We don't expect such (hypothetical) mechanism to have a large impact in the overall structure of the spacetime.

These geometries allow to quantify the effect of galaxies on GW emission, and our results lay to rest claims about strong suppression of GWs by the galactic potential~\cite{Kundu:1990ze} (see also Ref.~\cite{Price:1992pi} for an older argument that galactic potentials cannot indeed cause suppression).
A full understanding of the problem of GW emission and consequences for precision GW physics requires handling polar-type perturbations, which we leave for the future.

Finally, we note that there are other sources of dephasing of GWs, besides that studied here. Effectively, we studied here a pure spacetime effect, caused only by the different geometry.
Astrophysical binaries will also accrete and ``drag'' the surrounding medium, potentially leading to more significant effects, that add to the ones we described here~\cite{Eda:2013gg,Macedo:2013qea,Barausse:2014tra,Cardoso:2019rou,Annulli:2020lyc,Coogan:2021uqv,Derdzinski:2020wlw,Zwick:2021dlg}. Note, however, that such large dephasings due the dynamical friction require abnormally dense
dark matter spikes or the presence of accretion disks~\footnote{Dark matter spikes are expected to form close to compact objects~\cite{Gondolo:1999ef,Sadeghian:2013laa}, but are also easily suppressed 
by stars or BHs, or accretion by the central BH, induced by heating in its
vicinities~\cite{Merritt:2002vj,Bertone:2005hw,Merritt:2003qk}.}.
For a dark matter environment with density $\sim 10^3M_{\odot}/pc^3$, and a binary with parameters as in the discussion above ($M_{\rm BH}=10^6M_\odot, m_2=10M\odot$),
the dephasing due to dynamical friction is $\delta {\cal N}={\cal O}(1)$~\cite{Barausse:2014tra,Cardoso:2019rou}, therefore smaller than that due to the galactic geometry. The effect scales with the environmental density and is therefore larger for possible dark matter spikes, and in accretion disks. However, we note that we only computed the axial sector; the polar sector is underway and expected to lead to larger effects. Thus, the galactic geometry impact on the evolution of compact binaries must be considered.

\noindent{\bf{\em Acknowledgments.}}
We are indebted to José Natário for useful discussions and for pointing us to important literature regarding stationary configurations. We also thank Antoine Lehébel, Pedro Ferreira and David Pereñiguez for valuable discussions and feedback.
V. C. is a Villum Investigator supported by VILLUM FONDEN (grant no. 37766) and a DNRF Chair supported by the Danish National Research Foundation.
F.D. acknowledges financial support provided by FCT/Portugal through grant No. SFRH/BD/143657/2019. 
This project has received funding from the European Union's Horizon 2020 research and innovation programme under the Marie Sklodowska-Curie grant agreement No 101007855.
We thank FCT for financial support through Project~No.~UIDB/00099/2020.
We acknowledge financial support provided by FCT/Portugal through grants PTDC/MAT-APL/30043/2017 and PTDC/FIS-AST/7002/2020.
The authors would like to acknowledge networking support by the GWverse COST Action 
CA16104, ``Black holes, gravitational waves and fundamental physics.''
%

\bibliographystyle{h-physrev4}
\bibliography{references} 

\begin{thebibliography}{10}

\bibitem{Freese:2008cz}
K.~Freese,
\newblock EAS Publ. Ser. {\bf 36}, 113 (2009), [0812.4005].

\bibitem{Navarro:1995iw}
J.~F. Navarro, C.~S. Frenk and S.~D.~M. White,
\newblock Astrophys. J. {\bf 462}, 563 (1996), [astro-ph/9508025].

\bibitem{Clowe:2006eq}
D.~Clowe {\em et~al.},
\newblock Astrophys. J. Lett. {\bf 648}, L109 (2006), [astro-ph/0608407].

\bibitem{Bertone:2004pz}
G.~Bertone, D.~Hooper and J.~Silk,
\newblock Phys. Rept. {\bf 405}, 279 (2005), [hep-ph/0404175].

\bibitem{Kahlhoefer:2017dnp}
F.~Kahlhoefer,
\newblock Int. J. Mod. Phys. A {\bf 32}, 1730006 (2017), [1702.02430].

\bibitem{PerezdelosHeros:2020qyt}
C.~P\'erez de~los Heros,
\newblock Symmetry {\bf 12}, 1648 (2020), [2008.11561].

\bibitem{LIGOScientific:2016aoc}
LIGO Scientific, Virgo, B.~P. Abbott {\em et~al.},
\newblock Phys. Rev. Lett. {\bf 116}, 061102 (2016), [1602.03837].

\bibitem{Abbott:2020niy}
LIGO Scientific, Virgo, R.~Abbott {\em et~al.},
\newblock 2010.14527.

\bibitem{EventHorizonTelescope:2019dse}
Event Horizon Telescope, K.~Akiyama {\em et~al.},
\newblock Astrophys. J. Lett. {\bf 875}, L1 (2019), [1906.11238].

\bibitem{GRAVITY:2020gka}
GRAVITY, R.~Abuter {\em et~al.},
\newblock Astron. Astrophys. {\bf 636}, L5 (2020), [2004.07187].

\bibitem{Barack:2018yly}
L.~Barack {\em et~al.},
\newblock Class. Quant. Grav. {\bf 36}, 143001 (2019), [1806.05195].

\bibitem{Cardoso:2019rvt}
V.~Cardoso and P.~Pani,
\newblock Living Rev. Rel. {\bf 22}, 4 (2019), [1904.05363].

\bibitem{Bertone:2018krk}
G.~Bertone and T.~Tait, M.~P.,
\newblock Nature {\bf 562}, 51 (2018), [1810.01668].

\bibitem{Bar:2019pnz}
N.~Bar, K.~Blum, T.~Lacroix and P.~Panci,
\newblock JCAP {\bf 07}, 045 (2019), [1905.11745].

\bibitem{Brito:2015oca}
R.~Brito, V.~Cardoso and P.~Pani,
\newblock Lect. Notes Phys. {\bf 906}, pp.1 (2015), [1501.06570].

\bibitem{Sadeghian:2013laa}
L.~Sadeghian, F.~Ferrer and C.~M. Will,
\newblock Phys. Rev. D {\bf 88}, 063522 (2013), [1305.2619].

\bibitem{Eda:2013gg}
K.~Eda, Y.~Itoh, S.~Kuroyanagi and J.~Silk,
\newblock Phys. Rev. Lett. {\bf 110}, 221101 (2013), [1301.5971].

\bibitem{Macedo:2013qea}
C.~F.~B. Macedo, P.~Pani, V.~Cardoso and L.~C.~B. Crispino,
\newblock Astrophys. J. {\bf 774}, 48 (2013), [1302.2646].

\bibitem{Barausse:2014tra}
E.~Barausse, V.~Cardoso and P.~Pani,
\newblock Phys. Rev. D {\bf 89}, 104059 (2014), [1404.7149].

\bibitem{Baibhav:2019rsa}
V.~Baibhav {\em et~al.},
\newblock 1908.11390.

\bibitem{Seoane:2021kkk}
P.~A. Seoane {\em et~al.},
\newblock 2107.09665.

\bibitem{Cardoso:2019rou}
V.~Cardoso and A.~Maselli,
\newblock Astron. Astrophys. {\bf 644}, A147 (2020), [1909.05870].

\bibitem{Kavanagh:2020cfn}
B.~J. Kavanagh, D.~A. Nichols, G.~Bertone and D.~Gaggero,
\newblock Phys. Rev. D {\bf 102}, 083006 (2020), [2002.12811].

\bibitem{Tamanini:2019usx}
N.~Tamanini, A.~Klein, C.~Bonvin, E.~Barausse and C.~Caprini,
\newblock Phys. Rev. D {\bf 101}, 063002 (2020), [1907.02018].

\bibitem{1990ApJ...356..359H}
L.~{Hernquist},
\newblock The Astrophysical Journal {\bf 356}, 359 (1990).

\bibitem{1983MNRAS.202..995J}
W.~{Jaffe},
\newblock Monthly Notices of The Royal Astronomical Society {\bf 202}, 995
  (1983).

\bibitem{1962AJ.....67..471K}
I.~{King},
\newblock The Astrophysical Journal {\bf 67}, 471 (1962).

\bibitem{Gondolo:1999ef}
P.~Gondolo and J.~Silk,
\newblock Phys. Rev. Lett. {\bf 83}, 1719 (1999), [astro-ph/9906391].

\bibitem{1996ApJ...471...68C}
L.~{Ciotti},
\newblock The Astrophysical Journal {\bf 471}, 68 (1996), [astro-ph/9605084].

\bibitem{2004MNRAS.351...18B}
M.~{Baes} and H.~{Dejonghe},
\newblock Monthly Notices of the Royal Astronomical Society {\bf 351}, 18
  (2004), [astro-ph/0402527].

\bibitem{Einstein:1939ms}
A.~Einstein,
\newblock Annals Math. {\bf 40}, 922 (1939).

\bibitem{Geralico:2012jt}
A.~Geralico, F.~Pompi and R.~Ruffini,
\newblock Int. J. Mod. Phys. Conf. Ser. {\bf 12}, 146 (2012).

\bibitem{chandrasekhar1992mathematical}
S.~Chandrasekhar,
\newblock {\em The Mathematical Theory of Black Holes}International series of
  monographs on physics (Oxford University Press, 1992).

\bibitem{Cardoso:2008bp}
V.~Cardoso, A.~S. Miranda, E.~Berti, H.~Witek and V.~T. Zanchin,
\newblock Phys. Rev. D {\bf 79}, 064016 (2009), [0812.1806].

\bibitem{1911MNRAS..71..460P}
H.~C. {Plummer},
\newblock Monthly Notices of the Royal Astronomical Society {\bf 71}, 460
  (1911).

\bibitem{Liebling:2012fv}
S.~L. Liebling and C.~Palenzuela,
\newblock Living Rev. Rel. {\bf 15}, 6 (2012), [1202.5809].

\bibitem{Annulli:2020lyc}
L.~Annulli, V.~Cardoso and R.~Vicente,
\newblock Phys. Rev. D {\bf 102}, 063022 (2020), [2009.00012].

\bibitem{2018A&A...618L..10G}
{Gravity Collaboration} {\em et~al.},
\newblock Astronomy and Astrophysics {\bf 618}, L10 (2018), [1810.12641].

\bibitem{Maselli:2014fca}
A.~Maselli, L.~Gualtieri, P.~Pani, L.~Stella and V.~Ferrari,
\newblock Astrophys. J. {\bf 801}, 115 (2015), [1412.3473].

\bibitem{Regge:1957td}
T.~Regge and J.~A. Wheeler,
\newblock Phys. Rev. {\bf 108}, 1063 (1957).

\bibitem{Zerilli:1970wzz}
F.~J. Zerilli,
\newblock Phys. Rev. D {\bf 2}, 2141 (1970).

\bibitem{Maggiore:2007ulw}
M.~Maggiore,
\newblock {\em {Gravitational Waves. Vol. 1: Theory and Experiments}}Oxford
  Master Series in Physics (Oxford University Press, 2007).

\bibitem{1967ApJ...149..591T}
K.~S. {Thorne} and A.~{Campolattaro},
\newblock The Astrophysical Journal {\bf 149}, 591 (1967).

\bibitem{1983ApJS...53...73L}
L.~{Lindblom} and S.~L. {Detweiler},
\newblock The Astrophysical Journal {\bf 53}, 73 (1983).

\bibitem{Binnington:2009bb}
T.~Binnington and E.~Poisson,
\newblock Phys. Rev. D {\bf 80}, 084018 (2009), [0906.1366].

\bibitem{Damour:2009vw}
T.~Damour and A.~Nagar,
\newblock Phys. Rev. D {\bf 80}, 084035 (2009), [0906.0096].

\bibitem{Cardoso:2017cfl}
V.~Cardoso, E.~Franzin, A.~Maselli, P.~Pani and G.~Raposo,
\newblock Phys. Rev. D {\bf 95}, 084014 (2017), [1701.01116],
\newblock [Addendum: Phys.Rev.D 95, 089901 (2017)].

\bibitem{Cardoso:2019upw}
V.~Cardoso and F.~Duque,
\newblock Phys. Rev. D {\bf 101}, 064028 (2020), [1912.07616].

\bibitem{Berti:2009kk}
E.~Berti, V.~Cardoso and A.~O. Starinets,
\newblock Class. Quant. Grav. {\bf 26}, 163001 (2009), [0905.2975].

\bibitem{Jansen:2017oag}
A.~Jansen,
\newblock Eur. Phys. J. Plus {\bf 132}, 546 (2017), [1709.09178].

\bibitem{Cardoso:2017soq}
V.~Cardoso, J.~L. Costa, K.~Destounis, P.~Hintz and A.~Jansen,
\newblock Phys. Rev. Lett. {\bf 120}, 031103 (2018), [1711.10502].

\bibitem{Berti:2005ys}
E.~Berti, V.~Cardoso and C.~M. Will,
\newblock Phys. Rev. D {\bf 73}, 064030 (2006), [gr-qc/0512160].

\bibitem{Berti:2016lat}
E.~Berti, A.~Sesana, E.~Barausse, V.~Cardoso and K.~Belczynski,
\newblock Phys. Rev. Lett. {\bf 117}, 101102 (2016), [1605.09286].

\bibitem{Vishveshwara:1970zz}
C.~V. Vishveshwara,
\newblock Nature {\bf 227}, 936 (1970).

\bibitem{Cardoso:2016rao}
V.~Cardoso, E.~Franzin and P.~Pani,
\newblock Phys. Rev. Lett. {\bf 116}, 171101 (2016), [1602.07309],
\newblock [Erratum: Phys.Rev.Lett. 117, 089902 (2016)].

\bibitem{Cardoso:2016oxy}
V.~Cardoso, S.~Hopper, C.~F.~B. Macedo, C.~Palenzuela and P.~Pani,
\newblock Phys. Rev. D {\bf 94}, 084031 (2016), [1608.08637].

\bibitem{Price:1971fb}
R.~H. Price,
\newblock Phys. Rev. D {\bf 5}, 2419 (1972).

\bibitem{Ching:1994bd}
E.~S.~C. Ching, P.~T. Leung, W.~M. Suen and K.~Young,
\newblock Phys. Rev. Lett. {\bf 74}, 2414 (1995), [gr-qc/9410044].

\bibitem{Blanchet:1994ez}
L.~Blanchet and B.~S. Sathyaprakash,
\newblock Phys. Rev. Lett. {\bf 74}, 1067 (1995).

\bibitem{Cardoso:2017cqb}
V.~Cardoso and P.~Pani,
\newblock Nature Astron. {\bf 1}, 586 (2017), [1709.01525].

\bibitem{Zworski99}
M.~Zworski,
\newblock Notices Amer. Math. Soc. {\bf 46}, 319 (1999).

\bibitem{SjoZwo99}
J.~Sjöstrand and M.~Zworski,
\newblock Acta Mathematica {\bf 183}, 191  (1999).

\bibitem{Teukolsky:1974yv}
S.~A. Teukolsky and W.~H. Press,
\newblock Astrophys. J. {\bf 193}, 443 (1974).

\bibitem{LISA:2017pwj}
LISA, P.~Amaro-Seoane {\em et~al.},
\newblock 1702.00786.

\bibitem{Maggiore:2019uih}
M.~Maggiore {\em et~al.},
\newblock JCAP {\bf 03}, 050 (2020), [1912.02622].

\bibitem{Martel:2005ir}
K.~Martel and E.~Poisson,
\newblock Phys. Rev. D {\bf 71}, 104003 (2005), [gr-qc/0502028].

\bibitem{Maselli:2020zgv}
A.~Maselli, N.~Franchini, L.~Gualtieri and T.~P. Sotiriou,
\newblock Phys. Rev. Lett. {\bf 125}, 141101 (2020), [2004.11895].

\bibitem{Maselli:2021men}
A.~Maselli {\em et~al.},
\newblock 2106.11325.

\bibitem{Xu:2018wow}
Z.~Xu, X.~Hou, X.~Gong and J.~Wang,
\newblock JCAP {\bf 09}, 038 (2018), [1803.00767].

\bibitem{Cadoni:2017evg}
M.~Cadoni, R.~Casadio, A.~Giusti, W.~M\"uck and M.~Tuveri,
\newblock Phys. Lett. B {\bf 776}, 242 (2018), [1707.09945].

\bibitem{Buyle:2006vs}
P.~Buyle, E.~Van~Hese, S.~De~Rijcke and H.~Dejonghe,
\newblock Mon. Not. Roy. Astron. Soc. {\bf 375}, 1157 (2007),
  [astro-ph/0612307].

\bibitem{Chandrasekhar:1964zz}
S.~Chandrasekhar,
\newblock Astrophys. J. {\bf 140}, 417 (1964),
\newblock [Erratum: Astrophys.J. 140, 1342 (1964)].

\bibitem{1964ApJ...140..417C}
S.~{Chandrasekhar},
\newblock \apj {\bf 140}, 417 (1964).

\bibitem{1966ApJ...145..505B}
J.~M. {Bardeen}, K.~S. {Thorne} and D.~W. {Meltzer},
\newblock \apj {\bf 145}, 505 (1966).

\bibitem{Kokkotas:2000up}
K.~D. Kokkotas and J.~Ruoff,
\newblock Astron. Astrophys. {\bf 366}, 565 (2001), [gr-qc/0011093].

\bibitem{Dev:2003qd}
K.~Dev and M.~Gleiser,
\newblock Gen. Rel. Grav. {\bf 35}, 1435 (2003), [gr-qc/0303077].

\bibitem{Kundu:1990ze}
P.~K. Kundu,
\newblock Proc. Roy. Soc. Lond. A {\bf 431}, 337 (1990).

\bibitem{Price:1992pi}
R.~H. Price, J.~Pullin and P.~K. Kundu,
\newblock Phys. Rev. Lett. {\bf 70}, 1572 (1993), [astro-ph/9212005].

\bibitem{Coogan:2021uqv}
A.~Coogan, G.~Bertone, D.~Gaggero, B.~J. Kavanagh and D.~A. Nichols,
\newblock 2108.04154.

\bibitem{Derdzinski:2020wlw}
A.~Derdzinski, D.~D'Orazio, P.~Duffell, Z.~Haiman and A.~MacFadyen,
\newblock Mon. Not. Roy. Astron. Soc. {\bf 501}, 3540 (2021), [2005.11333].

\bibitem{Zwick:2021dlg}
L.~Zwick, A.~Derdzinski, M.~Garg, P.~R. Capelo and L.~Mayer,
\newblock 2110.09097.

\bibitem{Merritt:2002vj}
D.~Merritt, M.~Milosavljevic, L.~Verde and R.~Jimenez,
\newblock Phys. Rev. Lett. {\bf 88}, 191301 (2002), [astro-ph/0201376].

\bibitem{Bertone:2005hw}
G.~Bertone and D.~Merritt,
\newblock Phys. Rev. D {\bf 72}, 103502 (2005), [astro-ph/0501555].

\bibitem{Merritt:2003qk}
D.~Merritt,
\newblock Phys. Rev. Lett. {\bf 92}, 201304 (2004), [astro-ph/0311594].

\end{thebibliography}
\end{document}